  \documentstyle[12pt]{article}
  \def\bea{\begin{eqnarray}}
  \def\eea{\end{eqnarray}}
  \def\beq{\begin{equation}}
  \def\eeq{\end{equation}}
  
  \def\tF{{\tilde F}}
  \def\notD{\not{\hspace{-.07in}D}}
  \def\LLL{\sum_\Phi{\lambda_\Phi\Lambda_\Phi^2\over 32\pi^2}}
  \def\lll{\sum_P{\lambda_P\over 32\pi^2}}
  \def\ibar{\bar{\imath}}
  
  \def\ba{\bar{\alpha}}
  
  \def\bmu{\bar{\mu}}
  \def\[{\left [}
  \def\]{\right ]}
  \def\({\left (}
  \def\){\right )}

  \def\pp{\partial}
  \def\M{\bar{M}}
  
  \def\z{\bar{z}}

  \def\STr{{\rm STr}}
  \def\Tr{{\rm Tr}}
  \def\cA{{\cal A}}
  \def\K{{\cal K}}

  \def\f{\bar{f}}
  \def\F{{\cal F}}
  
  \def\hcA{\hat{{\cal A}}}
  
  \def\L{{\cal L}}
  \def\D{{\cal D}}
  \def\notcD{\not{\hspace{-.07in}\D}}
  \def\bl{\bar{\lambda}}
  \def\hf{\hat{\phi}}
  \def\hz{\hat{z}}

  \def\hA{\hat{A}}
  
  \def\W{\overline{W}}
  \def\hD{\hat{D}}
  
  \def\hV{\hat{V}}
  
  \def\m{\bar{m}}
  
  \def\cM{{\cal{M}}}
  
  \def\bc{\bar{\chi}}

  \def\A{\bar{A}}
  \def\bc{\bar{\chi}}
  \def\bps{\bar{\psi}}
  \def\Z{{\bar{Z}}}

  \def\bth{\bar{\theta}}
  \def\bv{\bar{\varphi}}
  
  \def\bF{\bar{F}}

\begin{document}

  \begin{titlepage}
  \begin{center}

  \hfill LBL-37425 \\
  \hfill UCB-PTH-95/19 \\
  \hfill June 1995 \\
  \hfill hep-th/9506149 \\[.3in]

  {\large \bf Pauli-Villars Regularization of Supergravity\\ and Field Theory
   Anomalies}\footnote{This work was supported in part by the Director,
  Office of High Energy and Nuclear Physics, Division of High Energy
  Physics of the U.S. Department of Energy under contract DE-AC03-76SF00098,
  in part by the National Science Foundation under grant PHY-90-21139.}
  \footnote{Talk given at SUSY95, l'\'Ecole Polytechnique, Palaiseau, France,
  May 15--19, 1995}\\[.2in]

  Mary K. Gaillard \\[.1in]

  {\em Department of Physics and Theoretical Physics Group, Lawrence Berkeley
  Laboratory,
  University of California, Berkeley, California 94720}\\[.5in]

  \end{center}
  \vskip 2in

  \begin{abstract}
  A procedure for Pauli-Villars regularization of locally and globally
  supersymmetric theories is described.  Implications for specific theories,
  especially those obtained from superstrings, are discussed with emphasis on
  the role of field theory anomalies.
  \end{abstract}
  \end{titlepage}

  \newpage

  Participation in this conference implies a belief that supersymmetry is
relevant
  to particle physics.  Since gravity is an established force of nature, we are
  inevitably led to take supergravity seriously as an effective,
nonrenormalizable
  theory valid in some energy range.  As such, we should be able to treat it at
  the quantum level, which means that we need a regularization procedure
  consistent with the symmetries of the theory.  In this talk I will describe a
  procedure~\cite{mk} for regulating the one-loop quadratic divergences of a
  general supergravity theory~\cite{crem}, with only the restriction that the
  gauge kinetic function is diagonal in gauge indices: $f_{ab}(Z) =
  \delta_{ab}k_af(Z)$.  In the case of global supersymmetry, the procedure has
  been generalized~\cite{mk2} to regulate all one-loop ultraviolet divergences
  for a general gauged nonlinear $\sigma$-model with no dilaton.  Following
some
  preliminaries to establish notation, specify the gauge fixing, {\it etc.},
  I will describe the Pauli-Villars field content and couplings.  Finally I
will
  mention several implications for and applications to physical issues that are
  especially relevant to effective supergravity theories from strings.

  The tree-level supergravity lagrangian~\cite{crem} I adopt, with $f(z) = x +
  iy$, is
  \bea {1\over\sqrt{g}}\L &=& {1\over 2}r - {x\over 4}F_{\mu\nu}F^{\mu\nu} -
  {y\over 4}{\tilde F}_{\mu\nu}F^{\mu\nu} + K_{i\m}\D^\mu z^i\D_\mu\z^{\m}- V
  \nonumber \\ & & + {ix\over 2}\bl\notD\lambda +
  iK^{i\m}\(\bc_L^{\m}\notD\chi_L^i + \bc_R^i\notD\chi_R^{\m}\)
  + e^{-K/2}\({1\over 4}f_i\A^i\bl_R\lambda_L - A_{ij}\bc^i_R\chi^j_L
  + {\rm h.c.}\)  \nonumber \\ & & +
  {1\over 4}\bps_\mu\gamma^\nu(i\notD + M)\gamma^\mu\psi_\nu
  - {1\over 4}\bps_\mu\gamma^\mu(i\notD + M)\gamma^\nu\psi_\nu \nonumber \\ & &
  -\[{x\over 8}\bps_\mu\sigma^{\nu\rho}\gamma^\mu\lambda_aF^a_{\nu\rho}
  + \bps_\mu\notcD \z^{\m}K_{i\m}\gamma^\mu L\chi^i
  {1\over 4}\bps_\mu\gamma^\mu\gamma_5\lambda^a\D_a
  + i\bps_\mu\gamma^\mu L\chi^im_i + {\rm h.c.}\]
  \nonumber \\ & & + \(i\bl^a_R\[2K_{i\m}(T_a\z)^{\m} -
  {1\over 2x}f_i\D_a - {1\over 4}\sigma_{\mu\nu}F^{\mu\nu}_af_i\]\chi^i_L +
  {\rm h.c.}\) + {\rm 4\;fermion \;terms}, \eea
  where
  \bea V &=& \hV + \D, \;\;\;\; \hV = e^{-K}(A_i\A^i - 3A\A), \quad
  \D = {1\over 2x}\D_a\D^a, \;\;\;\; \D_a = K_i(T^az^i),  \nonumber \\
  \M &=& (M)^{\dag} = e^{K/2}\(WR + \W L\), \;\;\;\; A = e^KW, \;\;\;\;
  \A = e^K\W, \;\;\;\; m_i = e^{-K/2}A_i. \eea
  $K(z,\z)$ is the K\"ahler potential, $W(z)$ is the superpotential, $T^a$ is a
  generator of the gauge group, and
  \beq A_{i_1\cdots i_n} = D_{i_1}\cdots D_{i_n}A, \;\;\;\; \A^{i_1\cdots i_n}
=
  D^{i_1}\cdots D^{i_n}\A, \qquad D^i = K^{i\m}D_{\m},\eeq
  with $D_i$ the scalar field reparameterization covariant derivative, and
  $K^{i\m}$ the inverse K\"ahler metric.
  The one-loop effective action is determined from the quadratic quantum
action:
  \bea &&\L_{quad}(\Phi,\Theta,c) = {1\over 2}\hf^I\hf^J\(\pp_I\pp_J +
  (A_I)^K_J\pp_K\)S + \L_{gf} + \L_{gh} =
  \nonumber \\ && \qquad -{1\over 2}\Phi^TZ^\Phi\(\hD^2_\Phi + H_\Phi\)\Phi
  + {1\over 2}\bar{\Theta}Z^\Theta\(i\notD_\Theta - M_\Theta\)\Theta
  + {1\over 2}\bar{c} Z^c\(\hD^2_c + H_c\)c + O(\psi_{cl}), \eea
  where $\phi^I = \Phi^I,\;\Theta^I,\; \pp_I = {\pp/\pp\phi^I},$
  and the column vectors,
  $$\Phi^T = (h_{\mu\nu},\hcA^a,\hz^i,\hz^{\m}),\;\;\;\;\;
  \Theta^T = (\psi_\mu,\lambda^a,\chi^i_L,\chi^{\m}_R,\alpha),\;\;\;
  \; c^T = (c_\nu, c^a, c_\alpha),$$
  represent the bose, fermion and ghost quantum degrees of freedom,
  respectively, with
  $ \alpha = -C\ba^T$ an auxiliary field introduced~\cite{us} to implement
  gravitino gauge fixing.  The connection $(A_I)^K_J$ is chosen so as to
preserve
  all bosonic symmetries, and also to simplify matrix elements involving the
  graviton. In particular the quantum variables
  $\hz^i,\hz^{\m}$ are normal coordinates in the space of scalar fields:
  $(A_i)^k_j = \Gamma^k_{ij},$ the affine connection associated with the
K\"ahler
  metric $K_{i\m}$, giving a scalar field reparameterization invariant
expansion.
  In (4) $\psi_{cl}$ represents background fermion fields that I set to zero;
that
  is, I calculate only the one-loop bosonic action.

  For the bose sector, I use a smeared gauge-fixing:
  \beq \L\to \L + \L_{gf}, \;\;\;\; \L_{gf} = -{\sqrt{g}\over
2}C_AZ^{AB}C_B,\;\;
  \;\; Z = \pmatrix{\delta^{ab}&0\cr0& -g^{\mu\nu}\cr}, \;\;\;\;
  C = \pmatrix{C_a\cr C_\mu\cr}. \eeq
  The Yang-Mills gauge-fixing term:
  $$ C^a = \D''^\mu\hcA^a_\mu + {i\over\sqrt{x}}K_{i\m}\[(T^a)^{\m}\hz^i
  - (T^a)^i\hz^{\m}\],$$
  preserves off-shell supersymmetry~\cite{barb} in the limit of global
  supersymmetry and coincides with the string-loop result~\cite{ant} for chiral
  multiplet wave function renormalization. The graviton gauge-fixing term:
  $$ \sqrt{2}C_\mu = \(\nabla^\nu h_{\mu\nu}
  - {1\over 2}\nabla_\mu h^\nu_\nu - 2\D_\mu z^IZ_{IJ}\hz^J +
  2\F^a_{\mu\nu}\hcA_a^\nu \), $$
  is the one originally introduced by 't Hooft and Veltman~\cite{tini},
  generalized~\cite{us} to include the Yang-Mills sector.
  The script quantum and classical Yang-Mills fields and field strengths are
  canonically normalized~\cite{noncan}:
  $$ \cA_\mu = \sqrt{x}A_\mu, \;\;\;\; \hcA_\mu = \sqrt{x}\hA_\mu, \;\;\;\;
  \F_{\mu\nu} = \sqrt{x}F_{\mu\nu}, \;\;\;\;
  \sqrt{x}\D_\mu A_\nu = \D'_\mu\cA_\mu,$$
  where $\D_\mu$ is the gauge and general covariant derivative, and
  $ \D'_\mu = \D_\mu - \pp_\mu x/ 2x,\;\;
  \D''_\mu = \D_\mu + \pp_\mu x/ 2x. $
  In the earlier literature two gravitino gauge fixing procedures have been
used:
  a) a Landau-type gauge~\cite{ino,josh} $\gamma\cdot\psi = 0$, implemented by
  the introduction of an auxiliary field, and b) a smeared gauge~\cite{lahanas}
  $\L\to\L - \bF\cM F,\; F = \gamma\cdot\psi, \; \cM = {1\over 4}\(i\notD
  + 2M_\psi\)$ supplemented with Nielsen-Kallosh ghosts.  Here I use an
unsmeared
  gauge $G = 0,$ with the gauge-fixing function~\cite{us}
  \beq
  G = -\gamma^\nu(i\notD - \M)\psi_\nu
  - 2(\notcD z^iK_{i\m}R\chi^{\m} + \notcD\z^{\m}K_{i\m}L\chi^i)
  + {x\over 2}\sigma^{\nu\rho}\lambda_aF^a_{\nu\rho}
  + 2im_I\chi^I - \gamma_5\D_a\lambda^a, \eeq
  where $D_\mu$ contains the spin and chiral K\"ahler connections.
  The quantum Lagrangian is obtained by the introduction of an auxiliary field
  $\alpha$: $\delta(G) = \int d\alpha\;{\rm exp}\(i\alpha G\), $
  and a shift in the gravitino field:
  $ \psi' = \psi +\gamma\alpha, \;\bps' = \bps + \ba\gamma, $
  so as to diagonalize the gravitino kinetic energy term.
  The ghost and ghostino determinants are obtained in the usual way as,
  respectively:
  $$ \(\hD^2_c + H_c\)^A_B = {\pp\over\pp\epsilon_A}\delta C_B,
  \;\;\;\; A,B = a,\mu,\;\;\;\;
  \(\hD^2_c + H_c\)^\alpha_\beta = {\pp\delta G^\alpha\over\pp\epsilon^\beta},
$$
  where $\hD_\mu$ is related to $\D_\mu$ or $D_\mu$ by additional connections.
  With these choices the one-loop bosonic action takes a very simple form:
  \bea S_1 &=& {i\over 2}\Tr\ln\(\hD_\Phi^2 + H_\Phi\)
  -{i\over 2}\Tr\ln\(-i\notD_\Theta + M_\Theta\)
  + {i\over2}\STr\ln\(\hD_c^2 + H_c\), \eea
  where
  \bea \STr\ln\(\hD_c^2 + H_c\) &=& 2\Tr\ln\(\hD_c^2 + H_c\)_{c_\alpha}
  - 2\Tr\ln\(\hD_c^2 + H_c\)_{c_{a,\mu}}, \nonumber \eea
  which just reduces to determinants of the form of those for
  scalars and spin-${1\over2}$ fermions. Moreover the ghost and non-ghost
  sectors have separately supersymmetric quantum spectra, except for the
  Yang-Mills fields:
  $$ {1\over 2}\(\Tr \;1\)_\Theta = \(\Tr \;1\)_\Phi -2N_G =
2N+2N_G+10.\;\;\;\;
  \(\Tr \;1\)_{c_\alpha} = 4, \;\;\;\; \(\Tr \;1\)_{c_{a,b}} = 4 + N_G, $$
  where $N (N_G)$ is the number of chiral (gauge) supermultiplets.
  To evaluate (7) I separate~\cite{josh,kamran} the fermion determinant into
  helicity-even and -odd parts:
  \bea
  -{i\over 2}\Tr\ln(-i\notD + M_\Theta) \equiv -{i\over 2}\Tr\ln\cM(\gamma_5)
   = T_- + T_+, \eea
  where here $D_\mu$ contains all fermion connections, and
  \bea T_- &=& -{i\over 4}\[\Tr\ln\cM(\gamma_5) - \Tr\ln\cM(-\gamma_5)\],\qquad
  T_+ = -{i\over 4}\[\Tr\ln\cM(\gamma_5) + \Tr\ln\cM(-\gamma_5)\], \nonumber \\
  \cM &=&\gamma_0(-i\notD + M_\Theta) = \pmatrix{\sigma_+^\mu D^+_\mu & M^+\cr
M^-
  &\sigma_-^\mu D^-_\mu\cr},\qquad
  \sigma_{\pm}^\mu = (1, \pm\vec \sigma). \eea
  Then defining
  $$ \hD^2_\Theta + H_\Theta \equiv
  \(-i\notD_\Theta + M_\Theta\)\(i\notD_\Theta + M_\Theta\), $$
  The one-loop bosonic action (7) reduces to:
  \bea S_1 &=& {i\over 2}\STr\ln\(\hD^2 + H\) + T_-. \nonumber \eea
  The helicity-odd term $T_-$ is at most logarithmically divergent, and is
  finite~\cite{us,kamran} in the absence of a dilaton, that is, for
  $f(Z) = g^{-2} + i\theta/8\pi^2 =$ constant.  It does not contribute to the
  effective actions considered in this talk.

  First consider the case $f(Z) = $ constant.  The quadratically divergent
  [$O(\Lambda^2)$] contributions to the one-loop bosonic action are:
  \bea \STr H_{grav} &=& -10V -2M_\psi^2 - {r\over2} - {x\over 2}F^2
  + 4K_{i\m}\D_\nu z^i\D^\mu\z^{\m} , \nonumber \\
  \STr H_\chi &=&  2N\(\hV + M_\psi^2 - {r\over4}\) + 2x^{-1}\D_aD_i(T^az)^i
  - 2R_{i\m}\(e^{-K}\A^iA^{\m} + \D_\nu z^i\D^\mu\z^{\m}\), \nonumber \\
  \STr H_{YM} &=&  2\D + {x\over 2}F^2 + N_G{r\over 2}. \eea
  First note the cancellation of the terms containing the squared Yang-Mills
field
  strength $F^2$. Assuming that the generators of the gauge group are
  traceless: Tr$T^a = 0$, the $z$-dependent terms can by regulated by the
  introduction  Pauli-Villars chiral supermultiplets:
  $Z_\alpha^I = (\Z_\alpha^{\bar{I}})^{\dag}, \; Z'^{\bar{I}}_\alpha =
  (\Z'^I_\alpha)^{\dag},\; \varphi^A = (\bv^A)^{\dag}$, where
  $Z^I$ transforms like $z^i$ under the gauge group, and $Z'^{\bar{I}}
  \sim\z^{\ibar}$  transforms according to the conjugate representation,
  with K\"ahler potential:
  \bea K(Z,\Z,\varphi,\bv) &=& \sum_{\alpha,I=i,M=m}K_{i\m}(z,\z)
  \(Z_\alpha^I\Z_\alpha^{\M} + \Z'^I_\alpha Z'^{\M}_\alpha\)
  + \sum_Ae^{\alpha_AK}\varphi^A\bv^A, \nonumber \eea
  superpotential:
  $$ W(Z,\varphi) = \sum_{A,I}\mu^\alpha_IZ^I_\alpha Z'^{\bar{I}}_\alpha
  + \sum_{A}\mu_A\(\varphi^A\)^2, $$
  and signature $\eta^{\alpha,A} = \pm 1$, which determines the sign of the
  contribution to the supertrace relative to that of a physical supermultiplet
  ({\it e.g.}, ghosts have signature $-1$).
  The Pauli-Villars contribution to $\STr H_\chi$ is:
  \bea \STr H^{PV}_\chi &=& 2\sum_{\alpha,A}\(2\eta_\alpha + \eta_A\)
  \(\hV + M_\psi^2 - {r\over4}\) \nonumber \\ & &
  + 4\sum_{\alpha,J}\eta_\alpha
  \[x^{-1}\D^aD_{(\alpha J)}(T_az)^{(\alpha J)} -
  R^{(\alpha J)}_{(\alpha J)i\m}\(\A^iA^{\m}e^{-K}
  + \D_\mu z^i\D^\mu\z^{\m}\)\] \nonumber \\ & &
  + 2\sum_A\eta_A\[x^{-1}\D^aD_A(T_az)^A
  - R^A_{A i\m}\(\A^iA^{\m}e^{-K} + \D_\mu z^i\D^\mu\z^{\m}\)\],\eea
  with:
  \bea \Gamma^A_{Bk} &=& \alpha_A\delta^A_BK_k \;\;\;\;
  R^A_{B k\m} = \alpha_A\delta^A_B K_{k\m}, \qquad
  \Gamma^{(I\alpha)}_{(J\beta),k} = \delta^\alpha_\beta\Gamma^i_{jk},
  \;\;\;\; R^{(I\alpha)}_{(J\beta)k\m} =
  \delta^\alpha_\beta \delta^I_JR^i_{jk\m},\nonumber \\
  D_A(T_az)^A &=& \alpha_A K_i(T_az)^i, \;\;\;\; D_I(T_az)^I = D_i(T_az)^i.
  \nonumber \eea
  To regulate the term proportional to the space-time curvature
  $r$, I add $U(1)$ gauge supermultiplets $W^a$ with signature
  $\eta^a$, chiral multiplets $e^{\theta^a} = \(e^{\bth^a}\)^{\dag}$
  with signature $\eta^a$, $U(1)_b$ charge $q_a\delta_{ab}$, and
  K\"ahler potential: $
  K(\theta,\bth) = {1\over2}\sum_a \nu_a e^{\alpha_a K}(\theta_a + \bth_a)^2$
  which is invariant under $U(1)_b$: $\delta_b\theta_a = -\delta_b\bth_a = iq_a
  \delta_{ab}$.  The  fields
  $(\theta^a + \bth^a)/\sqrt{2}$ combine with the $W^a$ to form vector
  supermultiplets with squared mass $\mu_a^2 =
(2x)^{-1}q^2_ae^{\alpha_aK}\nu_a$.
  The chiral supermultiplets contribute in the same way as $\varphi^A$ with
  $\alpha_A\to\alpha_a$; the vector supermultiplets contribute only to
  the $r$-term, with the opposite sign.  Then the
  overall contribution from light and heavy modes is:
  \bea
  \STr H' &=&  2\hV\[N\(1+2\sum_\alpha\eta_\alpha\) +
  \sum_A\eta_A\(1 - \alpha_A\)
  + \sum_a\eta_a\(1 - \alpha_a\) - 5\] \nonumber \\ & & +
  2M_\psi^2\[N\(1+2\sum_\alpha\eta_\alpha\) + \sum_A\eta_A\(1 - 3\alpha_A\)
  + \sum_a\eta_a\(1 - 3\alpha_a\) - 1\] \nonumber \\ & &
  - {r\over2}\[N\(1+2\sum_\alpha\eta_\alpha\) + \sum_A\eta_A + 1 - N_G\]
  \nonumber \\ & &
  + 2\(1+2\sum_\alpha\eta_\alpha\)\[{1\over x}\D_a(T^az)^i
  - R_{i\m}\(e^{-K}\A^iA^{\m} + \D_\mu z^i\D^\mu\z^{\m}\)\]
  \nonumber \\ & & + 2\(K_{i\m}\D_\mu z^i\D^\mu\z^{\m} - 2\D\)\(2 -
  \sum_A\eta_A\alpha_A - \sum_a\eta_a\alpha_a\). \eea
  The finiteness condition $\STr H' = 0$ requires:
  \bea 0 = 1+2\sum_\alpha\eta_\alpha = \sum_A\eta_A
  + \sum_a\eta_a - 7 = \sum_A\eta_A + 1 - N_G = 2 -
  \sum_A\eta_A\alpha_A - \sum_a\eta_a\alpha_a.\eea
  Note that there are only four independent conditions possible, but these are
  sufficient to cancel eight {\it a priori} independent quadratically divergent
  contributions.
  Next, taking, for example, $q_a = 1,\;\mu_I^\alpha =
  \beta_\alpha^Z\mu_I,\; \mu_A = \beta_A\mu_\varphi, \;
  \nu_a = x\beta^2_a|\mu_\theta|^2,$
  cancellation of the $O(\mu^2\ln\Lambda^2)$ terms requires the additional
  conditions: $\sum_\alpha\eta_\alpha^Z\(\beta^Z_\alpha\)^2 =
  \sum_A\eta_A\(\beta_A\)^2 = \sum_a\eta_a\(\beta_a\)^2 = 0.$ The
  result for the $O(\mu^2)$ part of $S_0 + S_1 = \int
   d^4x\(\L_0 + \L_1\)$ is:
  \bea \L_0(g_{\mu\nu}^0,K) + \L_1 &=& \L_0(g_{\mu\nu},K + \delta K),
  \qquad g_{\mu\nu} = g_{\mu\nu}^0\(1 + \epsilon\), \nonumber \eea \bea
  \epsilon &=& -\lll e^{-K}A_{PQ}\A^{PQ}, \qquad
  \delta K =\lll\(e^{-K}A_{PQ}\A^{PQ} -4\K_P^P\), \eea
  where $P,Q=$ represent all heavy fields, and~\cite{sigma}
  $\lambda_P = 2{\displaystyle{\sum_p}}\eta^P_p\(\beta^P_p\)^2\ln\beta^p_P.$
  This result can be expressed in terms of effective cut-offs $\Lambda_\Phi$:
  \bea \epsilon &=& \LLL\zeta'_\Phi, \qquad
  \delta K = \LLL\zeta_\Phi, \eea
  where:
  \bea\lambda_\Phi &=&
2\sum_p\eta^\Phi_p\(\beta^\Phi_p\)^2\ln\beta^p_\Phi,\qquad
  \zeta_Z=\zeta_\varphi=\zeta'_Z =\zeta'_\varphi=1,\quad\zeta_\theta =
  -4\;\zeta_\theta' = 0,\nonumber \\
  \Lambda^2_{Z^I} &=& \sum_{m=M}e^K\(K^{i\m}\)^2\mu_I\bmu_{\M}, \qquad
  \Lambda^2_\varphi = e^{K(1-2\alpha_\varphi)}|\mu_\varphi|^2, \qquad
  \Lambda^2_\theta = |\mu_\theta|^2e^{\alpha_\theta K}. \nonumber \eea
  Note however that if there are three or more terms in the sum over $p$, the
  sign of $\lambda_\Phi$ is indeterminate~\cite{sigma}, so caution should be
  used in making conclusions about the implications of these terms for the
  stability of the effective potential.

  Before including the dilaton, I note that there is an ambiguity in the
  separation (9) of the fermion determinant into helicity-even and -odd
  contributions. For example, if
  $D_\mu = \pp_\mu + V_\mu + iA_\mu\gamma_5,\; M = m + m'\gamma_5$,
  we would identify $D_\mu^{\pm} = \pp_\mu + V_\mu \pm iA_\mu,\; M^{\pm} = m
  \mp m'.$  However terms even and odd in $\gamma_5$ can be interchanged by the
  use of the identities:
  $$\gamma_5  = (i/24)\epsilon^{\mu\nu\rho\sigma}\gamma_\mu
  \gamma_\nu\gamma_\rho\gamma_\sigma, \;\;\;\;
  \sigma_{\mu\nu} = i\gamma_5\sigma_{\mu\nu}, \;\;\;\; etc.$$
  The choice is generally dictated by gauge or K\"ahler covariance.  However
there
  is an off-diagonal gaugino-$\alpha$ mass term:
  \bea M_{\alpha\lambda^a} &=& -\sqrt{x\over2}F_a^{\mu\nu}\sigma_{\mu\nu}
  = -\sqrt{x\over2}\(\alpha F_a^{\mu\nu} + i\beta\gamma_5
   \tF_a^{\mu\nu}\)\sigma_{\mu\nu},\qquad
  \alpha + \beta = 1.\eea
  The one-loop action
  $S_1$ is invariant under the choice of $\alpha$ only if the integrals are
  finite.  For an arbitrary choice we obtain, instead of (10)
  \bea \STr H_{grav} &\ni& {x\over 2}F^2(\alpha^2 - \beta^2 -2), \;\;\;
  \STr H_{YM} \ni  {x\over 2}F^2(\alpha^2 - \beta^2).\nonumber \eea
  The choice $\alpha=1$ is the ``supersymmetric'' one, in that it matches
  analogous matrix elements in bose and ghost sectors, resulting in the
  cancellation of the $F^2$ terms in (10).  We now include a dilaton, that is
we
  take $f_{ab} = \delta_{ab}f, \;\; f= x+iy\ne$ constant (which is trivially
  generalized to $f_{ab} = \delta_{ab}k_af,\;k_a=$ constant, and so includes
all
  known string models).  There is a dilatino-gaugino mass term:
  \bea M_{\chi^i\lambda^a} &=& -i{f_i\over4\sqrt{x}}\(\gamma F_a^{\mu\nu}
  + i\delta\gamma_5\tF_a^{\mu\nu}\)\sigma_{\mu\nu},\qquad
  \gamma + \delta = 1, \;\;\;\; f_i = \pp_if, \nonumber \eea
  and an additional gaugino connection: \bea
  A^\mu_{\lambda^a\lambda^b} &=& -\delta_{ab}{\pp^\mu
y\over2x}\(i\epsilon\gamma_5
  - \zeta{\epsilon^{\lambda\nu\rho\sigma}\over24}\gamma_\lambda
  \gamma_\nu\gamma_\rho\gamma_\sigma\), \qquad
   \epsilon + \zeta = 1,\nonumber  \eea
  that give the additional contributions to the supertraces:
  \bea
  \STr H_{YM} &\ni&  {f_i\f^i\over 4x}F^2(\gamma^2 - \delta^2) -
  N_G\(2M_\lambda^2
  + {1\over2x^2}\[\pp_\mu x\pp^\mu x + (3 - 2\zeta^2)\pp_\mu y\pp^\mu y\]\),
  \nonumber \\ \STr H_{grav} &\ni& {f_i\f^i\over 4x}F^2(\gamma^2 - \delta^2)
   - {f_i\f^i\over2x^2}\D,
  \qquad \STr H_\chi \ni {f_i\f^i\over2x^2}\D, \eea
  where $M_\lambda^2 = (2x)^{-2}e^{-K}f_i\f^jA_j\A^i,\; \f^i = K^{i\m}\f_{\m}.$
  In this case, the ``supersymmetric'' choices that match matrix
  elements in the bosonic and ghost sectors are
  $\gamma=\delta={1\over2},\;\epsilon=0,\;\zeta=1$, and the $F^2$ terms again
  cancel; the remaining terms:
  \bea \STr H \ni - N_G\(2M_\lambda^2
  + {1\over2x^2}\[\pp_\mu x\pp^\mu x + \pp_\mu y\pp^\mu y\]\),\eea
  can be regulated by the introduction of additional Pauli-Villars chiral
  multiplets:
  $\pi^\alpha = \(\bar{\pi}^\alpha\)^{\dag}$  with
  $$ K(\pi,\bar{\pi}) = \sum_\alpha(f + \f)|\pi^\alpha|^2, \;\;\;\;
  W(\pi) = \sum_\alpha\beta_\pi^\alpha \mu_\pi(z)(\pi^\alpha)^2,
  \;\;\;\; \eta^\pi_\alpha = \pm 1.$$
  Finiteness requires:
  $\sum_\alpha\eta^\pi_\alpha = + N_G, \;\;
  \sum_\alpha\eta_\alpha^\pi\(\beta_\pi^\alpha\)^2 = 0,$ and the
  results (12--15) are modified accordingly; the sums in (14,15) are extended
to
  include, respectively $P= \pi_\alpha,\;\Phi= \pi$, with:
  $$ \Lambda^2_\pi = {e^K\over4x^2} |\beta_\alpha^\pi\mu_\pi|^2
  \delta_\beta^\alpha\(\beta_\alpha^\pi\)^2 , \qquad \zeta_\pi =
  \zeta'_\pi = 1.$$
  The result $\zeta = 1$ has implications for chiral/linear multiplet duality.
  In this case the $y$-axion contribution to the gaugino connection is
  $A_\mu =
2xh^{\nu\rho\sigma}\gamma_{[\mu}\gamma_\nu\gamma_\rho\gamma_{\sigma]}
  ,$ where $h^{\nu\rho\sigma}= \epsilon^{\nu\rho\sigma\mu}\pp_\mu y/4x^2$ is
the
  three-form that is dual to $y$-axion in the absence of interactions; there is
  a similar term in the vector connection. This suggests that the linear
  supermultiplet formalism~\cite{counter,linear} is the natural framework for
  describing the dilaton supermultiplet.  It has been shown~\cite{frad} for
  several models that axion/three-form duality holds at the quantum level up to
  finite topological anomalies.  In the presence of fermion couplings to the
  dilaton, there are additional
  anomalies; interchanging terms in $T_+$ and $T_-$ is analogous to
  shifting the integration variable. The linearly divergent triangle diagram
  leads to an ill-defined finite chiral anomaly that is fixed by imposing local
  gauge invariance; in the present case supersymmetry resolves the ambiguity.
  In addition, with the choice $\zeta=1$, the dilaton supermultiplet
contributes
  a purely
  ``vector-like'' gaugino connection, and there is no $y$-axion analogue of
  the modular anomaly (the Im$tF\tF$ term induced at one loop).  This result
  coincides with the conclusion of~\cite{tom}, where it was argued that a
  $yF\tF$ term is inconsistent with the linearity constraint of the linear
  multiplet formulation.

  To fully regulate the theory, including all logarithmic divergences, requires
  additional Pauli-Villars fields and/or couplings.  For example,
  to regulate the Yang-Mills sector we must include in the set $\varphi^A$
$N_G$
  chiral multiplets $\varphi^A_a = (\bv^A_a)^{\dag}$ that transform according
to
  the adjoint representation of the gauge group, with
  $ \sum_{A,a}\eta^a_A = 3N_G$. In this case the
  effective cut-off has been determined~\cite{tom} by imposing supersymmetric
  matching of chiral and conformal anomalies, giving $\alpha_A^a = {1\over 3}$.
  Imposing the full finiteness conditions on
  $\STr H'^2$ may constrain the other parameters $\alpha_A,\alpha_a,
  \alpha_\alpha.$  This program has been carried out~\cite{mk2} only in the
case
  of global supersymmetry with no dilaton. In this case only Pauli-Villars
  chiral multiplets
  are needed.  The full superpotential and K\"ahler potential are,
respectively:
  \bea W(z,Z_\alpha,Z'_\alpha,\varphi_\beta) &=& W(z) +
  \sum_{\alpha,I}\mu^\alpha_IZ^I_\alpha Z'^I_\alpha
  + \sum_{a,\beta}\mu_\beta\varphi_a^\beta\varphi^a_\beta
  + {1\over 2}\sum_\alpha a^\alpha W(Z)_{ij}Z^I_\alpha Z^J_\alpha \nonumber \\
& &
  + g^2\sum_\gamma b^\gamma\varphi^a_\gamma(Z'_\gamma T_aZ),\nonumber \\
  K(z^i,\z^{\ibar},\phi_{PV}) &=& K(z^i,\z^{\ibar}) +
  \sum_\beta \bv^\beta_a\varphi^a_\beta + \sum_\alpha K_{i\m} \nonumber \\ & &
  (z,\z)\(Z_\alpha^I\Z_\alpha^{\M} + \Z'^I_\alpha Z'^{\M}_\alpha\)
  + \sum_{\alpha;\;I,J=i,j}\beta_\alpha K_{ij}(z,\z)Z_\alpha^IZ^J_\alpha
  + O\(\phi_{PV}^2\),\nonumber \eea
  with
  $$ \alpha = 1,\cdots, N_I, \;\;\;\; \beta = 1,\cdots, N_\varphi, \;\;\;\;
  \gamma = 1,\cdots, N_{I\varphi}, \;\;\;\; N_{I\varphi} \le {\rm min}\{
  N_I,N_\varphi\}, $$
  and additional constraints on $a^\alpha,b^\gamma,\beta_\alpha$ are imposed by
  the finiteness requirement.

  Some of the cut-offs have a straightforward physical interpretation in the
  context of string theory. For chiral matter in torus compactifications or the
  untwisted sector of orbifold compactifications, the cut-off (in reduced
Planck
  units and assuming a common $vev$ for the moduli),
  $ \left. \Lambda_{Z^I}\right|_{\rm vac} = \Lambda_{\rm comp} = 1/R, $
  is just the inverse radius of compactification. The cut-off for the
  Yang-Mills sector,
  $\Lambda_{\rm gauge} = g^{-{2\over3}}\Lambda_{\rm comp},$
  can be interpreted in terms of the same cut-off, but incorporating the 2-loop
  correction to the $\beta$-function, which assures the supersymmetric relation
  between chiral and conformal anomalies.  For twisted chiral matter, the
  interpretation is less transparent: $\left.
  \Lambda^I_{\rm twisted}\right|_{\rm vac} =
  \Lambda_{\rm comp}\(g^{-4}\Lambda^4_{\rm comp}\)^{3q_I -1}. $
  The twisted sector must be included to assure both the cancellation of the
  modular anomaly and consistency with string loop
calculations~\cite{dkl,ant2}.
  Including the Green-Schwarz counterterm~\cite{counter}, matching field theory
  anomalies--as calculated with the above prescriptions--to string loop
  calculations, and using the all-loop supersymmetric renormalization group
  invariant function of~\cite{russians}, it was shown~\cite{tom} that the
  two-loop gauge unification scale (as conventionally defined by
  phenomenologists) is the
  string scale.  The same conclusion has been reached in~\cite{kl} where the
  same RGE invariant function, but a different calculational procedure, was
used.

  The extension of the procedure described here to a full regularization
  of supergravity is in progress~\cite{rey}.  Two cases are of special
interest.
  For a compact $\sigma$-model coupled to gravity, the classical ratio
  $f_\pi/m_{Pl}$ is fixed by the Bagger-Witten quantization
condition~\cite{bag},
  and one may ask whether this condition
  persists at the quantum level, which would require $\delta K = 0$ in (15).
  Since (in a class of $\sigma$-models) the scalar Ricci tensor
  $R_{i\m}$ is proportional to the K\"ahler metric $K_{i\m}$, the cut-offs in
(15)
  can be chosen such that the quadratically divergent parts of gravity and
  chiral loops cancel in $\delta K.$ The possibility of a full cancellation of
  the ultraviolet divergences as well as of the chiral anomaly is under
  investigation. More interesting for string theory are the noncompact
  $\sigma$-models that arise in torus and orbifold compactifications.  These
  models possess classical noncompact, nonlinear symmetries that contain a
  discrete subgroup of modular transformations.  The same considerations hold
at
  the one-loop level as for the compact case.  The possibility of a
regularization
  procedure that respects the full continuous symmetry as well as the discrete
  modular symmetry has potentially important implications for phenomenology.
  The parameters introduced above to specify the couplings of the PV sector are
  in general field-dependent: $\mu = \mu(z)$, $\nu = \nu(z,\z)$.  For example,
in
  superstring theory there is an invariance under a modular transformation:
  $K\to K+ F(z) + \bF(\z),\;\;W\to e^{-F(z)}W,$ that is
  unbroken by perturbative quantum corrections.  Thus
  $Z^I_\alpha$ has the same modular weight as $z^i$ and
  $\varphi^A$ has modular weight $-\alpha_A/2$; the $z$-dependence of
  $\mu(z)$, $\nu_a(z,\z)$ must be chosen accordingly; this field dependence can
be
  interpreted as threshold effects from integration over heavy modes.
  Typically, one expects: $\mu(z) \propto \eta(T)^p$, where $T$ is a modulus
  and $\eta(T)$ is the Dedekind $\eta$-function. Such terms break the
  continuous classical symmetry, thereby destroying the
  no-scale structure of these models ($\langle V\rangle = 0$) and
  the protection of a mass hierarchy $\Delta m_{SUSY}\ll m_{\tilde G}$
  that is desirable both for phenomenology and cosmology, as has been discussed
in
  several lectures at this conference.  These problems might
  be avoided if an anomaly free regularization is possible.

  Full Pauli-Villars regularization of supergravity including the dilaton can
be
  investigated once the ultraviolet divergences have been fully determined for
  this case~\cite{kamran}.

  \vskip .2in
  \noindent{\bf Acknowledgements.} This work was supported in part by the
  Director, Office of Energy Research, Office of High Energy and Nuclear
Physics,
  Division of High Energy Physics of the U.S. Department of Energy under
Contract
  DE-AC03-76SF00098 and in part by the National Science Foundation under grant
  PHY-90-21139.

  \end{document}